\documentclass[fleqn,usenatbib]{mnras}
\usepackage[T1]{fontenc}
\usepackage{ae,aecompl}
\usepackage{graphicx}
\usepackage{newtxtext,newtxmath}
\usepackage{color}
\usepackage{ulem}
\newcommand{\Nbody}{$N$-body }
\newcommand{\MMi}{M/M_{\rm i}}
\newcommand{\rj}{r_{\rm J}}
\newcommand{\rhrj}{r_{\rm h}/r_{\rm J}}

\newcommand{\rlrj}{r_{\rm L}/r_{\rm J}}
\newcommand{\trh}{t_{\rm rh,i}}
\newcommand{\Rh}{R_{\rm h}}
\newcommand{\vphi}{v_{\rm \phi}}
\newcommand{\vpk}{v_{\rm pk}}
\newcommand{\rpk}{R_{\rm pk}}
\newcommand{\vpksigmao}{v_{\rm pk}/\sigma_0}
\newcommand{\vphisigma}{v_{\rm \phi}/\sigma}
\newcommand{\vphisigmap}{(v_{\rm \phi}/\sigma)_{\rm pk}}
\newcommand{\vphisigmao}{v_{\rm \phi}/\sigma_0}
\newcommand{\oO}{\omega_{\rm in}/\Omega}
\newcommand{\ooutO}{\omega_{\rm out}/\Omega}
\newcommand{\oout}{\omega_{\rm out}}

\title[Tidally limited clusters rotational properties]{Kinematical evolution of tidally limited star clusters: rotational properties}

\author[M. Tiongco, E. Vesperini and A.~L. Varri]{
Maria A. Tiongco,$^{1}$\thanks{E-mail: mtiongco@indiana.edu}
Enrico Vesperini,$^{1}$
and Anna Lisa Varri$^{2}$
\\
$^{1}$Department of Astronomy, Indiana University, Bloomington, IN 47405, USA\\
$^{2}$Institute for Astronomy, University of Edinburgh, Royal Observatory, Blackford Hill, Edinburgh EH9 3HJ, UK
}

\date{Accepted XXX. Received YYY; in original form ZZZ}

\pubyear{2016}

\begin{document}
\label{firstpage}
\pagerange{\pageref{firstpage}--\pageref{lastpage}}
\maketitle

\begin{abstract}
We present the results of a set of N-body simulations following the long-term evolution of the rotational properties of star cluster models evolving in the external tidal field of their host galaxy, after an initial phase of violent relaxation. The effects of two-body relaxation and escape of stars lead to a redistribution of the ordered kinetic energy from the inner to the outer regions, ultimately determining a progressive general loss of angular momentum; these effects are reflected in the overall decline the rotation curve as the cluster evolves and loses stars.

We show that all of our models share the same dependence of the remaining fraction of the initial rotation on the fraction of the initial mass lost. As the cluster evolves and loses part of its initial angular momentum, it becomes increasingly dominated by random motions, but even after several tens of relaxation times, and losing a significant fraction of its initial mass, a cluster can still be characterized by a non-negligible ratio of the rotational velocity to the velocity dispersion. This result is in qualitative agreement with the recently observed kinematical complexity which characterizes several Galactic globular clusters.
\end{abstract}


\section{Introduction}

\begin{figure*}
	\includegraphics{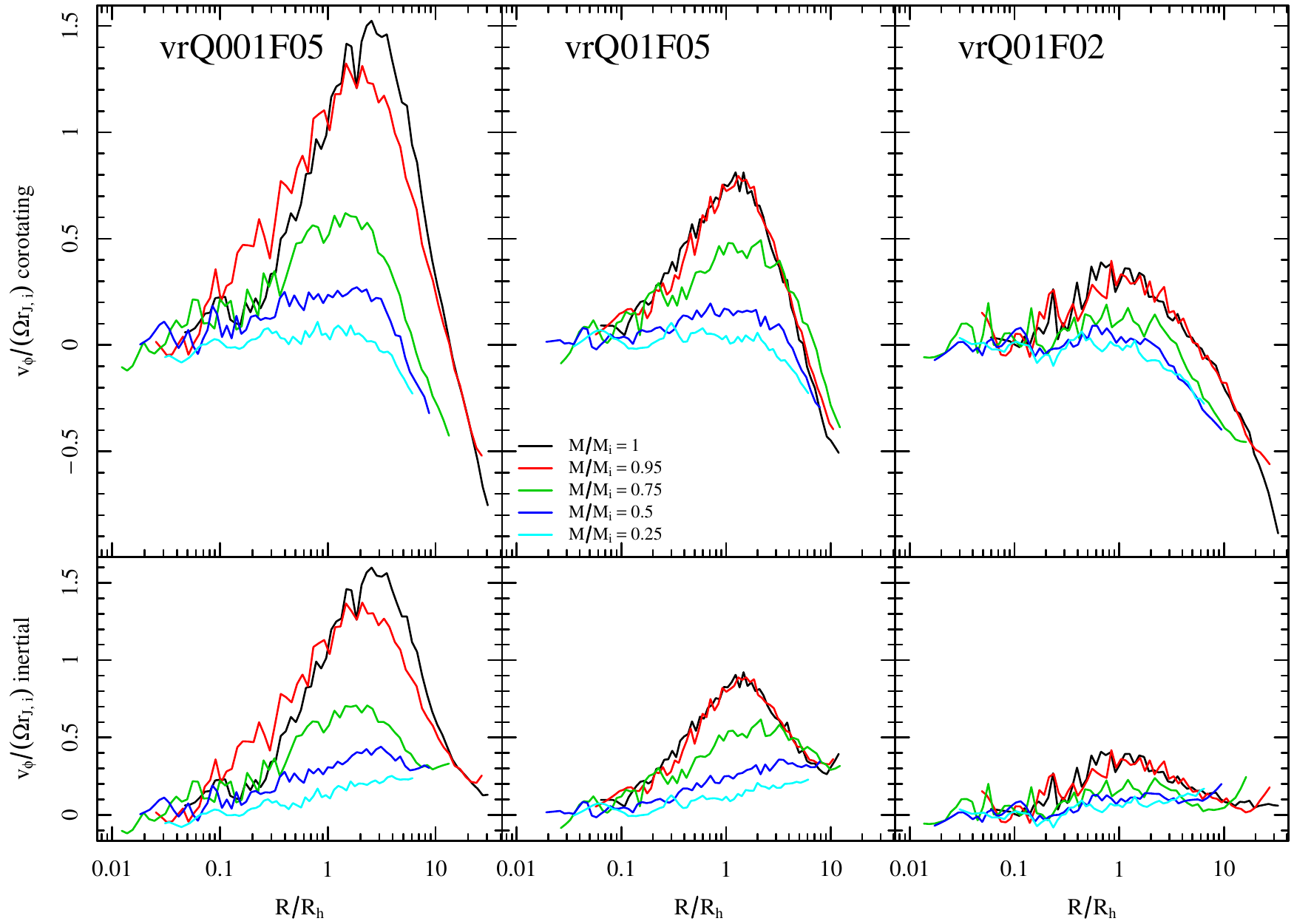}
    \caption{Evolution of the rotation curves for all the models.  Each coloured line represents the system different times chosen when the system contains a certain fraction of its initial mass, $\MMi$. In the construction of the profiles all particles within the Jacobi radius have been considered.  Top panels: The $\vphi$ component of the velocity normalized to the rotation speed of the coordinate system at the initial Jacobi radius, $\Omega r_{\rm J,i}$ in the corotating reference frame.  Bottom panels: The same profiles in the non-rotating, i.e. inertial reference frame.}
    \label{fig:vphievol}
\end{figure*}

\begin{figure*}
	\includegraphics{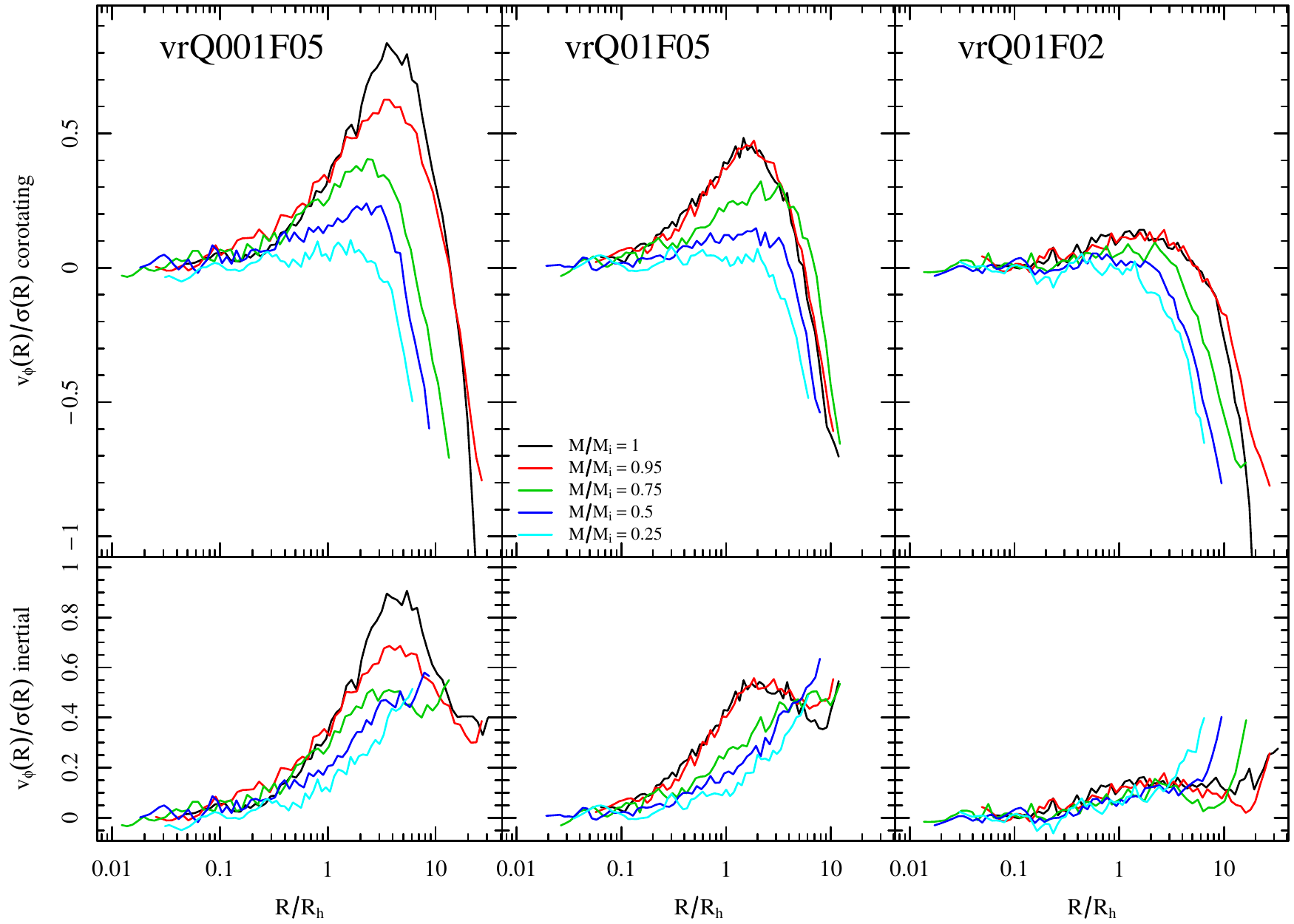}
    \caption{Time evolution of the rotational velocity profile divided by the velocity dispersion profile.  See Fig. \ref{fig:vphievol} for further details about the adopted colour scheme and organization of the panels.}
    \label{fig:vsigrevol}
\end{figure*}

\begin{figure*}
	\includegraphics{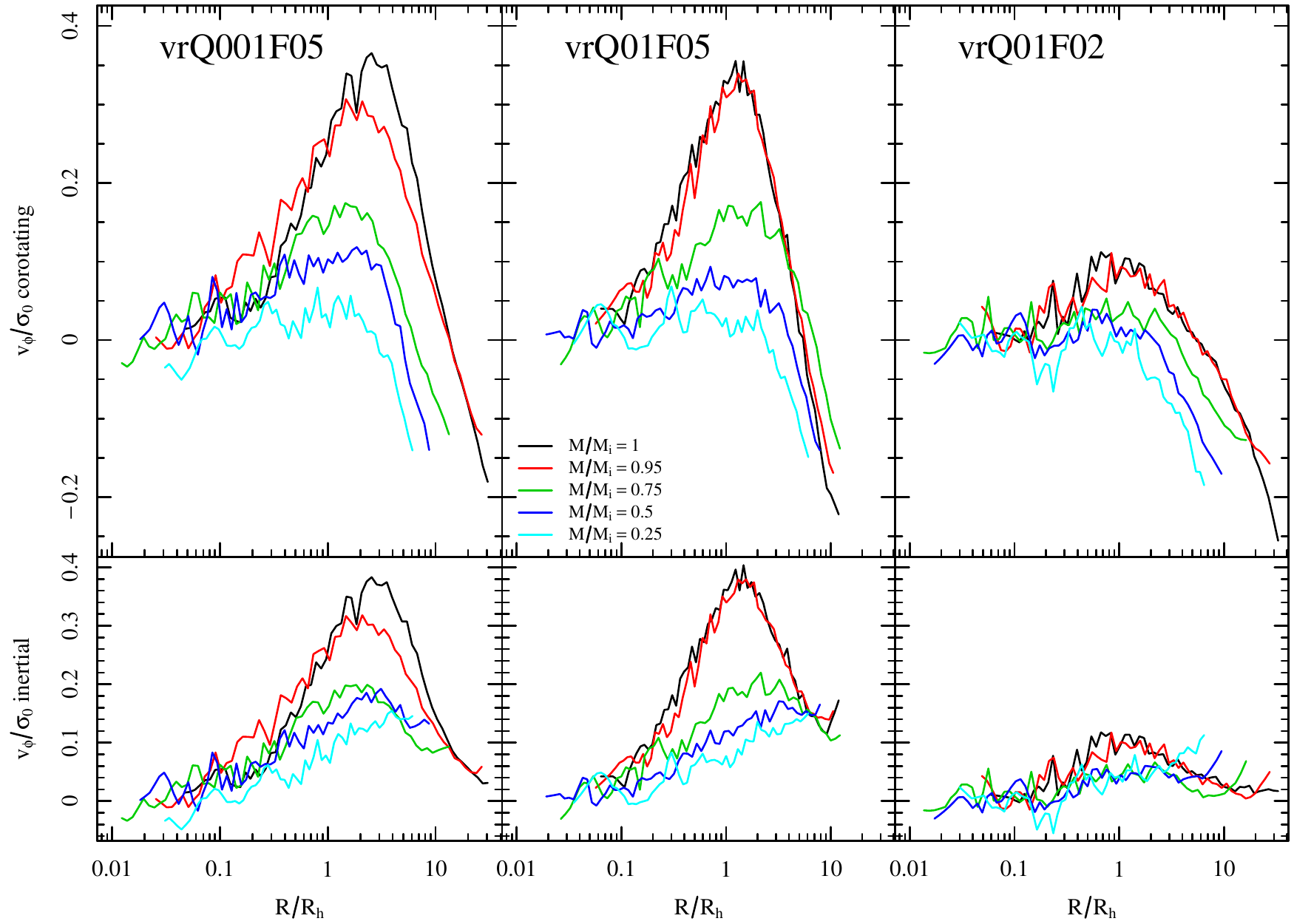}
    \caption{Time evolution of the rotational velocity profile divided by the central velocity dispersion.  See Fig. \ref{fig:vphievol} for further details about the adopted colour scheme and organization of the panels.}
    \label{fig:vsig0evol}
\end{figure*}

The traditional picture of globular clusters is that they are spherically symmetric, isotropic, and non-rotating stellar systems, a picture long supported by the success of spherical and isotropic equilibria such as \citet{king1966}  and \citet{wilson1975} models to fit the cluster surface density profiles \citep[e.g.][]{mclaughlin2005,miocchi2013}. However, observational studies aimed at exploring the morphological and the kinematical properties of globular clusters are revealing a more complex picture showing deviations from spherical symmetry (see e.g. \citealt{chen2010}; see also the early pioneering study of \citealt{white1987}), and an internal kinematics characterized by the presence of radial anisotropy \citep[see e.g.][]{richer2013,watkins2015,bellini2015}, and non-negligible rotation (see e.g. \citealt{meylan1997}, and more recent works including \citealt{lane2011}, \citealt{bellazzini2012}, \citealt{fabricius2014}, \citealt{kacharov2014}, \citealt{lardo2015}, \citealt{cordero2017}).

Observational studies have shown that the values of  cluster rotational velocities normalized to the cluster central velocity dispersions  range from about 0.05 to about 0.3 \citep[see e.g.][]{bellazzini2012,fabricius2014,kacharov2014}. Accurate radial profile measurements of the rotational velocities are in general not available for globular clusters, and the actual peak in the radial profile $V_{\rm rot}(r)/\sigma(r)$ could be larger than the values reported in  existing observational studies, which are often based on measurements averaging the rotational velocities over a broad range of cluster-centric distances. Moreover line-of-sight observations provide information on projected velocities and do not allow a direct determination of the actual $V_{\rm rot}/\sigma$.

It is also important to emphasize that the present-day rotation in globular clusters is likely to be the remnant of a stronger early rotation weakened by the effects of two-body relaxation (see \citealt{mapelli2017} for an example of hydrodynamical simulations producing rotating young massive clusters and \citealt{henaultbrunet2012} for measurements of rotation in young massive cluster R136); specifically, during the cluster long-term evolution, the magnitude of the rotation decreases due to escaping stars carrying away angular momentum, and the angular momentum of the remaining stars being redistributed on the relaxation timescale \citep[see e.g.][]{akiyama1989,einsel1999,kim2002,kim2004,kim2008,ernst2007,hong2013}.  The results of these pioneering theoretical works have shown that rotation can affect several important aspects of a cluster evolution such as core collapse and the mass loss rate. 

The neglect of rotation in  the study of globular cluster dynamics does not appear to be justified either by theoretical or observational studies, and as we enter the \textit{Gaia} era and new results from ESO/VLT and HST studies are sharpening the dynamical picture of clusters, a renewed effort to further our understanding of the role of internal rotation and its evolution in globular clusters is necessary.

In this paper, we present the results of a set of \Nbody simulations following the long-term evolution of tidally-limited models of star clusters, as initially resulting from an early phase of violent relaxation in the presence of an external tidal field. The focus of this study is on the evolution of the angular momentum content, by means of a complete characterization of the rotational properties of the systems under consideration.  We will devote particular attention to the dynamical interplay between the internal rotation and the effects of the tidal field of the host galaxy, with the goal of investigating the connection between the evolution of the angular momentum and the different stages of cluster evolution.

The outline of the paper is as follows: in Section 2, we describe our method and the initial conditions adopted. We present our results in Section 3 with a general discussion of the evolution of the rotation curve along with a discussion focused on some specific features such as the peak  of the rotation curve and the strength of rotation in the cluster inner and outermost regions. Finally, in Section 4 we summarize our conclusions.

\section{Method and Initial Conditions}
\label{sec:methods}

\begin{table}
\caption{Summary of simulations}
\label{tab:details}
\begin{tabular}{@{}cccc}
\hline
Model ID & 
$N_{\rm i,avg}$ &
$(\rhrj)_{\rm i,avg}$ & 
realizations\\
\hline
vrQ01F05 & 14 860 & 0.095 & 4\\
vrQ01F02 & 16 287 & 0.036 & 4\\
vrQ001F05 & 17 091 & 0.036 & 4\\
\hline
\end{tabular}
\end{table}

The \Nbody models featured in this paper were originally presented in \citet{tiongco2016a}, in the context of a study of the long-term evolution of the velocity anisotropy of tidally limited star clusters.  The simulations follow the evolution of systems that first undergo violent relaxation in the host galaxy tidal field, modelled as a point-mass.  The clusters are assumed to be on circular orbits, and equations of motions are solved in a reference frame corotating with the cluster around the host galaxy centre \citep[see e.g.][]{heggie2003}, with angular velocity $\Omega$.  Particles are removed from the simulation once they move beyond two times the Jacobi radius, $\rj$.  We carry out our simulations using the {\sevensize NBODY6} code \citep{aarseth2003} accelerated by a GPU \citep{nitadori2012} on the {\sevensize BIG RED II} cluster at Indiana University.

For the initial violent relaxation phase we use different initial values of the virial ratio, $Q=T/|V|$ (where T and V are the system kinetic energy and potential energy respectively), and different ratios of the initial cluster limiting radius (the radius measuring the total initial size of the cluster) to the Jacobi radius (the limit on the cluster size imposed by the external tidal field of the cluster host galaxy; see e.g. \citealt{heggie2003}), $\rlrj$.  We use the following parameters for the initial conditions of the violent relaxation phase: (Model ID, $Q$, $\rlrj$): (vrQ01F05, 0.1, 0.5), (vrQ01F02, 0.1, 0.2), and (vrQ001F05, 0.01, 0.5).  The systems are initially spherical,  homogeneous, and have equal-mass particles, beginning with a slightly larger number of particles in order to take into account the loss of stars in the violent relaxation phase.

\citet{vesperini2014} characterized in detail the kinematical properties that emerge from the violent relaxation process under the influence of the tidal field and the effects of the Coriolis force on the stars' orbits. We refer to that paper for further details on the early evolution of kinematical properties; for the purposes of this paper, the main kinematical feature of these models is that they are characterized by differential rotation around an axis perpendicular to the orbital plane. 
In the frame of reference corotating with the cluster, the rotation is prograde in the inner regions and retrograde in the cluster outermost regions; more specifically, the rotation curve increases from the cluster centre, peaks at a distance from the cluster centre equal to approximately the half-mass radius, and then decreases to a retrograde rotation in the outermost regions. The black lines in the top panels of Fig. \ref{fig:vphievol} show the rotational velocity radial profiles emerging at the end of the violent relaxation phase as measured in the co-rotating frame of reference (the rotational velocity measured in the non-rotating coordinate system is shown in the bottom panels of Fig \ref{fig:vphievol}).

 In this paper, we focus on the subsequent \textit{long-term} evolution of the rotational properties of these systems.  From here onward, we mark the time at the end of the violent relaxation phase (when the large scale structural oscillations associated with the initial virialization process end, typically after a few dynamical times; see \citealt{vesperini2014}) as our initial time, $t = 0$.  We measure the essential properties of the systems (such as the number of particles and the ratio of the half-mass radius to the Jacobi radius) at this time  and we refer to these properties as the initial conditions of the long-term evolution. We list  these properties in Table \ref{tab:details}, and note that for each model,  these values are the averages obtained from four different simulations following the evolution of different random realizations of the same model. When we show the time evolution of the cluster rotational properties, time is normalized to the cluster initial relaxation time $t_{\rm rh,i}=0.138 N^{1/2}r_{\rm h}^{3/2}/(\langle m\rangle^{1/2} G^{1/2} \log(0.11 N))$ where $N$ is the number of stars, $\langle m \rangle $ is the star mean mass, and $r_{\rm h}$ is the 3D half-mass radius (see e.g. \citealt{spitzer1987}). For a cluster with, for example, $\langle m \rangle =0.6 M_{\odot}$ and $N=2\times 10^5$ and $r_{\rm h}=3$ pc, $t_{\rm rh,i}\sim 600 $ Myr. All the simulations are run until 75\% of the mass they had at the end of violent relaxation is lost.

\section{Results}

\subsection{Evolution of the rotation curve}
\label{sec:profevol}

We start our analysis by considering the time evolution of the rotation curves of our three models.  In Figs \ref{fig:vphievol}-\ref{fig:vsig0evol}, we plot the rotational velocity ($\vphi$, as measured about the $z$-axis) for each of the models versus the projected radius $R$ normalized to the projected half-mass radius, $\Rh$. We normalize $\vphi$ in a few different ways: to the rotation speed of the coordinate system at the initial Jacobi radius ($\Omega r_{\rm J,i}$, Fig. \ref{fig:vphievol}), to the averaged velocity dispersion $\sigma = \sqrt{\frac{1}{3}(\sigma_{\rm r}^2 + \sigma_{\rm \phi}^2 + \sigma_{\rm \theta}^2)}$ in cylindrical bins measured at $R$ (Fig. \ref{fig:vsigrevol}), and to the central velocity dispersion as sometimes done in observational studies (Fig. \ref{fig:vsig0evol}).  We define the central velocity dispersion as the velocity dispersion of the innermost 500 stars in the system.

The profiles are constructed by combining 5 snapshots around the desired time, taking into consideration all particles within the Jacobi radius and  binning them in cylindrical shells with heights encompassing the entire cluster.  All quantities desired are measured in each bin, and this process is repeated for each of the four realizations.  Then at each radius $R$, the median value of all the realizations is taken to create the profiles shown in Figs \ref{fig:vphievol}-\ref{fig:vsig0evol}.

The strength of the initial rotation in the three models is determined by the initial conditions of the violent relaxation phase: systems with a colder initial virial ratio or higher initial $\rlrj$ develop stronger rotation (see \citealt{vesperini2014} for further details).  
The three models, although characterized by different strengths of the initial rotation, share a number of common features in the evolutionary path of their rotational velocity radial profiles. Specifically:
\begin{enumerate}
\item For all models rotation becomes weaker as the system evolves and loses mass; as pointed out in the Introduction this implies that the rotation observed in clusters today is just the remnant of a stronger initial rotation and  a weak present-day rotation can not be used to rule out  the importance of rotation for the cluster dynamical history.
\item A non-negligible rotation is retained for a large fraction of the cluster life; for example the vrQ001F05 model is still characterized by a peak $\vphisigma$ equal to about 0.35 after losing 25\% of its initial mass (see Fig. \ref{fig:vsigrevol}, top left panel) and the model vrQ01F05 has a peak $\vphisigma$ equal to about 0.3 after losing 25\% of its initial mass (see Fig. \ref{fig:vsigrevol}, top middle panel).
\item The value of the peak in the profiles of $\vphisigma$ and $\vphisigmao$ decreases as the cluster evolves and loses mass, but its location within the cluster is approximately constant during the entire cluster evolution and found at about $1-2 \Rh$.
\item All models are initially characterized by an inner prograde rotation and outer retrograde rotation. As the cluster evolves, the inner prograde rotation is gradually erased. The outer retrograde rotation emerging at the end of the initial violent relaxation, as a result of the symmetry breaking introduced by the external tidal field in the collapse phase and of the effects of the Coriolis force on the orbit of the stars \citep[see][]{vesperini2014} persists for all models for the entire evolution of the cluster, due to the subsequent interplay between the evolution of the fraction of prograde and retrograde stars and the effects of the tidal field \citep[see][]{tiongco2016b}.
\end{enumerate}
In the next sections, we discuss in further detail the evolution of the main features of the rotation curves of our models.

\subsection{Evolution of the peak of the rotation curve}
\label{sec:peakevol}

\begin{figure}
	\includegraphics[width=3.3in,height=3.3in]{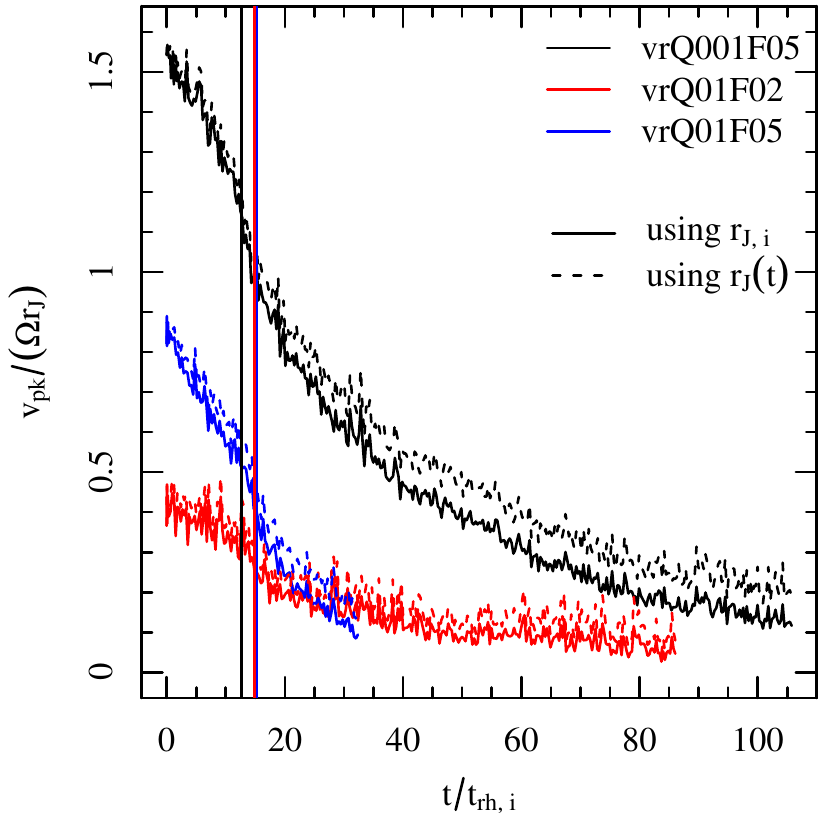}
    \caption{Evolution of the magnitude of the peak of the rotation curve for all the models as a function of time.  The two differing line types distinguish between normalizing with the initial Jacobi radius or the one at that time. The vertical lines mark the time of core collapse, calculated as median of the values obtained in the different model realizations.}
    \label{fig:vpkt}
\end{figure}

\begin{figure}
	\includegraphics[width=3.3in,height=3.3in]{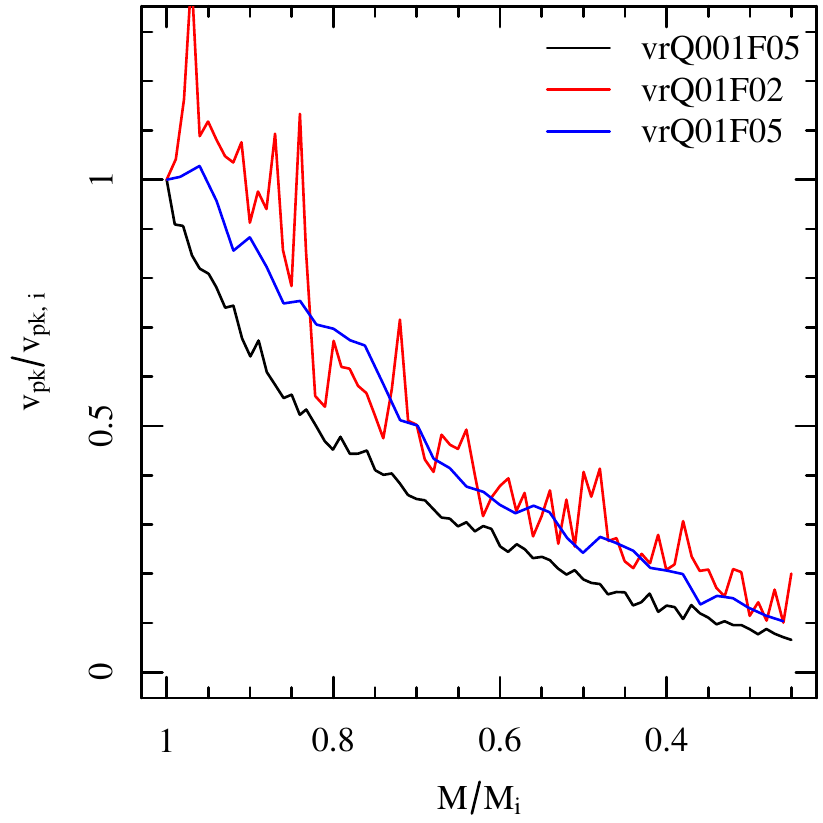}
    \caption{Evolution of the magnitude of the peak of the rotation curve normalized to its initial value as a function of fraction of initial mass remaining in the cluster.}
    \label{fig:vpkm}
\end{figure}

\begin{figure}
	\includegraphics[width=3.3in,height=3.3in]{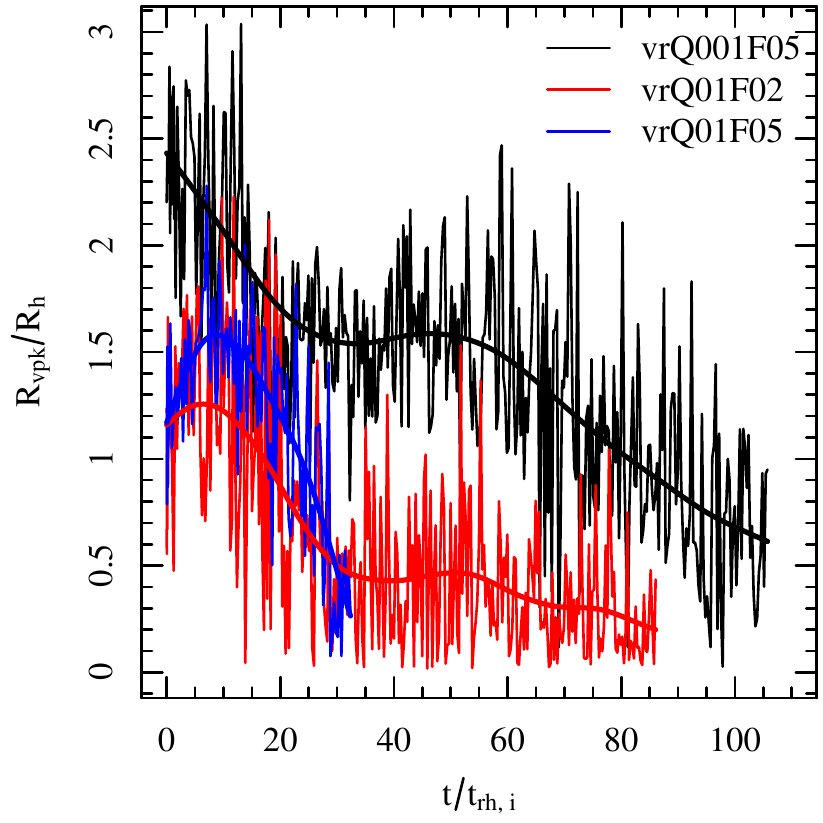}
    \caption{Evolution of the location of the peak of the rotation curve for all the models as a function of time.  The thicker lines are a smoothing spline fit added to guide the eye.}
    \label{fig:rpkt}
\end{figure}

Though in the previous section we characterized the evolution of rotation throughout the system, observationally such a complete characterization of the rotational velocity profile is extremely difficult.  A well-established method to measure the rotation in a cluster is through spectroscopy and line-of-sight radial velocity measurements of a sample of stars; the samples in current studies can reach up to a few hundred stars and although they rarely span a range of radial distances broad enough to construct a complete rotation curve (see e.g. \citealt{bianchini2013}, \citealt{kacharov2014}, and \citealt{cordero2017} for some recent studies of the rotational properties of selected Galactic globular clusters), they can be sufficient to identify the peak of the rotation curve. 

In the previous section, we have shown that the magnitude of the peak of the rotation curve decreases with time and that its distance from the cluster centre relative to the cluster half-mass radius remains approximately constant around 1-2 $\Rh$.
In Fig. \ref{fig:vpkt}, we take a closer look at the time evolution of the magnitude of the peak of the rotation curve,  $\vpk$.  Fig. \ref{fig:vpkt} shows that although $\vpk$ immediately starts to decreases, its rate of decrease undergoes a transition at the time of core collapse when $\vpk$ starts to decrease significantly more rapidly; we interpret this transition in the time evolution of $\vpk$ as the consequence of the onset of the post-core collapse expansion. The subsequent evolution is characterized by a gradual slowdown in the decrease rate of $\vpk$ allowing the cluster to retain a non-negligible value of $\vpk$ after several tens of initial half-mass relaxation times.

In  Fig. \ref{fig:vpkm}, we show the evolution of $\vpk$, this time normalized to the initial value of $\vpk$, as a function of the fraction of the initial mass remaining in a cluster; as discussed in the Introduction, the combined effect of angular momentum transport from the inner to the outer regions and star escape results in the overall loss of angular momentum, and thus contributes to the decline in the strength of rotation. Fig. \ref{fig:vpkm} shows that the systems we have studied almost universally lose the same fraction of their initial rotation for the same amount of mass lost, and even after losing about 50-60\% of their initial mass, they still retain some of their initial rotation.

As for the location of the peak, $\rpk$, Fig. \ref{fig:rpkt} shows that the peak location relative to the half-mass radius begins near 1-2 $\Rh$ for all of the models, then moves slightly inward over time.  This likely reflects the post core-collapse expansion of the half-mass radius, while the cluster itself is decreasing in size as it loses stars and the Jacobi radius moves inward.

\begin{figure}
	\includegraphics[width=3.3in,height=3.3in]{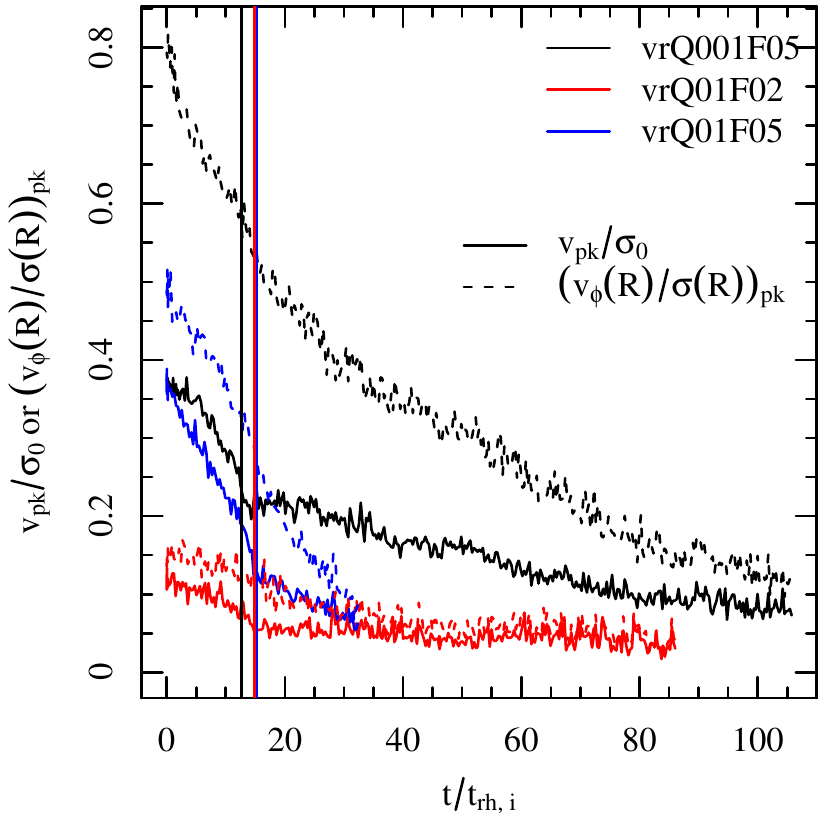}
    \caption{Evolution of the magnitude of the peak of the rotation curve normalized to the velocity dispersion for all the models as a function of time.  The two differing line types distinguish between $\vpksigmao$ and the peak of the $\vphisigma$ profile.  The vertical lines mark the time of core collapse.}
    \label{fig:vspkt}
\end{figure}

\begin{figure}
	\includegraphics[width=3.3in,height=3.3in]{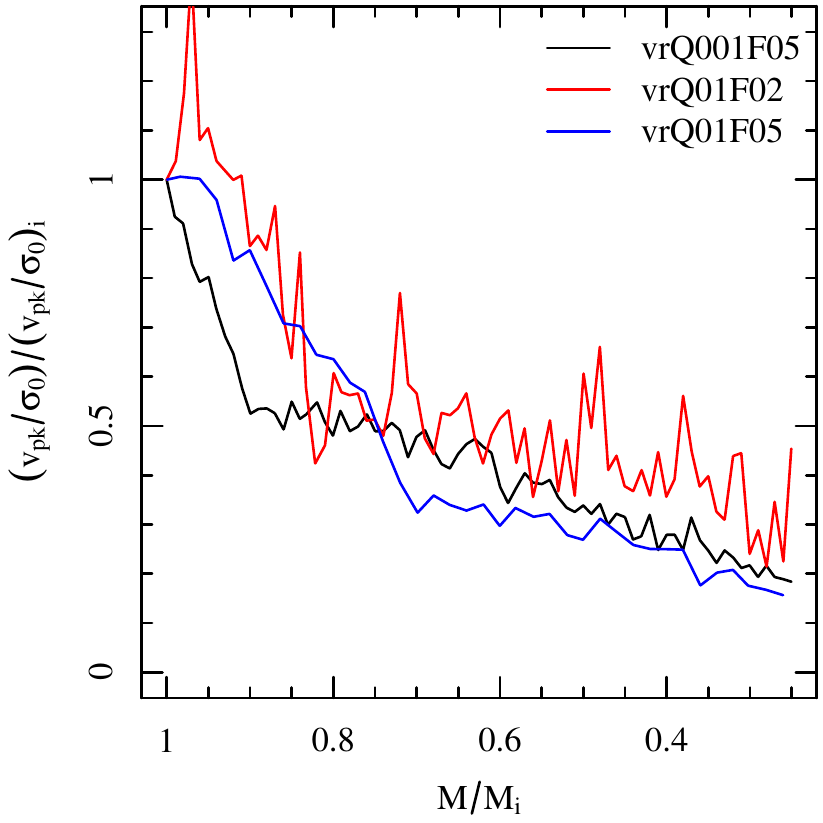}
    \caption{Evolution of the magnitude of $\vpksigmao$, normalized to its initial value as a function of the fraction of the initial mass remaining in the cluster.}
    \label{fig:vs0pkm}
\end{figure}

Next, we look at the evolution of $\vpksigmao$ as a function of time (Fig. \ref{fig:vspkt}, where we also show the evolution of the peak of the $\vphi(R)/\sigma(R)$ profile), and  normalized to its initial value versus the fraction of initial mass remaining (Fig. \ref{fig:vs0pkm}). This normalization using the central velocity dispersion is often adopted in the literature because of the lack of observational measurements of the full radial profile of the velocity dispersion.

Fig. \ref{fig:vspkt} clearly illustrates how the ratio of the rotational velocity to the velocity dispersion, and therefore the estimate of the relative importance of ordered and random motions in a cluster, depends on which value of the velocity dispersion is used for the normalization. It is interesting to point out how these figures show that clusters can retain values of $\vpksigmao$ similar to those found in observational studies  (see the initial discussion in the Introduction) after several tens of initial relaxation times and after having lost a significant fraction of their initial mass.  Fig. \ref{fig:vs0pkm} shows a similar trend to Fig. \ref{fig:vpkm} where the systems almost universally lose the same fraction of their initial $\vpksigmao$ for the same amount of mass lost (we also observe a similar trend for the fraction of initial peak value of $\vphi(R)/\sigma(R)$ versus mass lost).

The time evolution in Fig. \ref{fig:vspkt} is due to the combined effect  of the evolution of $\vphi$ and $\sigma$; it is interesting to point out that, similarly to what was already found in Fig. \ref{fig:vpkt}, the rate of evolution of $\vpksigmao$ appears to undergo a sharp transition to slower rates after core collapse (at about 15 $\trh$); we interpret this as the result of the decrease in velocity dispersion associated with the post-core collapse expansion \citep[see e.g.][]{giersz1994}.

\begin{figure*}
	\includegraphics{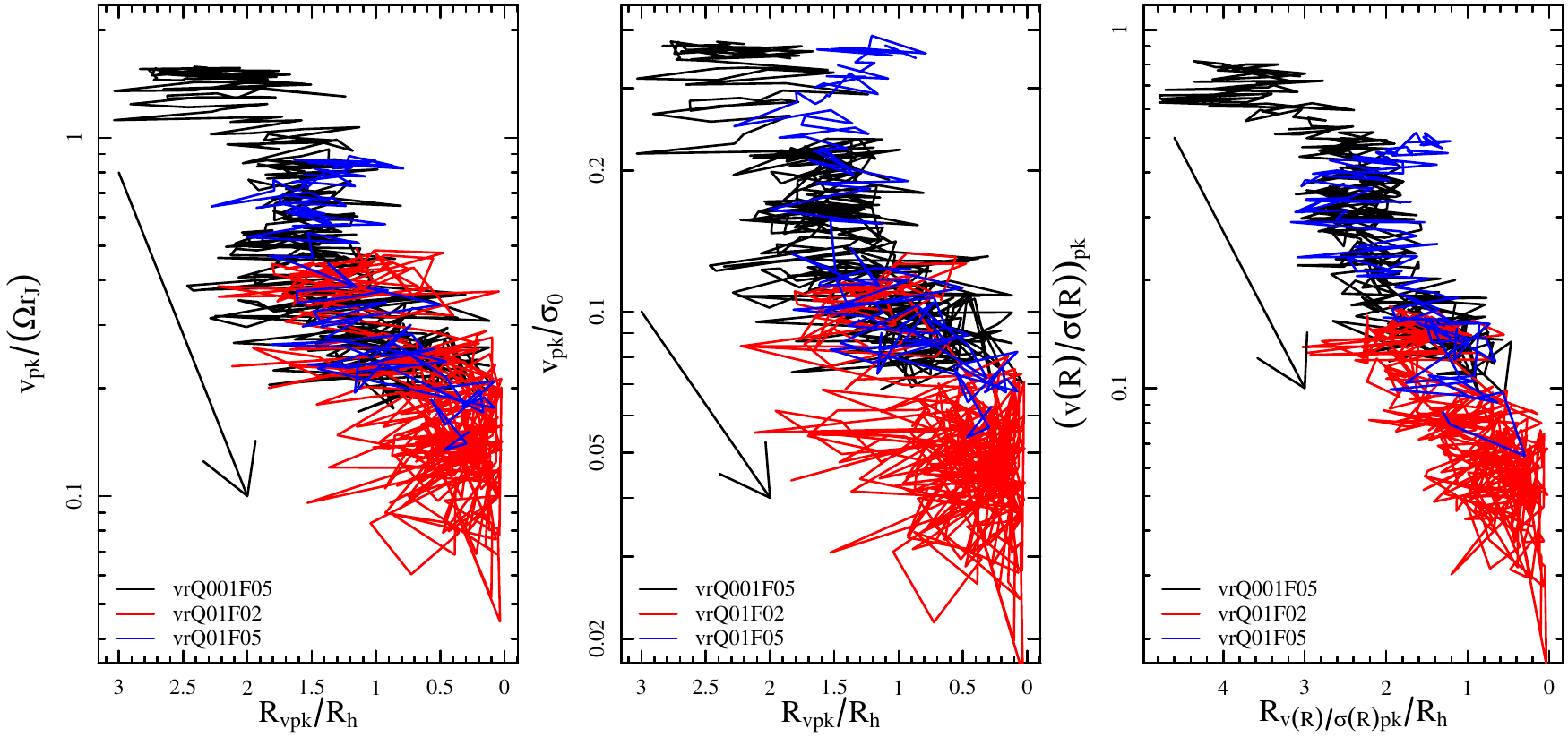}
    \caption{Left: Evolution of the magnitude of the peak of the rotation curve, normalized to $\Omega$ times the current Jacobi radius, as a function of its location at that time.  Middle: Evolution of $\vpksigmao$ as a function of the its location at that time.  Right: Evolution of the peak of the $\vphisigma$ profile as a function of its location at that time.  The arrows show the direction of evolution.}
    \label{fig:vpkrpk}
\end{figure*}

Finally, in Fig \ref{fig:vpkrpk} we show the evolution of the magnitude of the rotation peak in the different forms calculated above as a function of its location within the cluster.  The models appear to fall along a universal line (although with some significant noise) and follow a trend which further illustrates the evolution towards smaller values of the rotation peak at radii slightly closer to the cluster centre.

\subsection{Evolution of the slope of the rotation curve in the inner regions}
\label{sec:innerevol}

A number of observational studies have provided a measure of the rotation of the cluster inner regions. In particular, in a recent study  \citet{fabricius2014} have used data acquired with an Integral Field Unit spectrograph to measure rotation in the central regions of 11 Galactic globular clusters.  Although these data provide only an approximate global measurement of the rotational properties of the central and intermediate regions of the clusters and do not shed light on the details of the rotational velocity radial profile, they are nevertheless an important addition to the study of cluster kinematics and have provided further evidence of the presence of non-negligible rotation in globular clusters, even when they are in a relatively advanced stage of their dynamical evolution. 

In this section, we focus our attention on the cluster inner rotation; in order to quantify the strength of a cluster inner rotation we have measured the slope of the inner rotation curve by using both a linear fit of the inner rotation curve and the method employed by \citet{fabricius2014}.  The latter method involves fitting a two-dimensional function to the radial velocity field of the cluster inner regions: the gradient of this function provides a measure of the slope of the inner rotation which, in the case of solid-body rotation, coincides with the cluster angular velocity.  We refer to the measured slope as $\omega_{\rm in}$ and normalize it to the angular velocity of the cluster orbital motion around the host galaxy, $\Omega$.  For our calculation we have used particles between 0.1-0.5$\Rh$, combining 5 snapshots around the desired time with, typically, $\sim$4000 particles per snapshot at the beginning and $\sim$1000 particles per snapshot at the end of the simulation, and for the calculation based on the method adopted in \citet{fabricius2014} we have used the radial velocities measured along a line of sight perpendicular to the rotation axis.

Figs \ref{fig:oOtrhi} and \ref{fig:ominomini} show the evolution of $\oO$, found using the linear fitting method, respectively as a function of the time normalized to the initial half-mass relaxation time, and then normalized to its initial value as a function of the fraction of the initial mass remaining in the cluster.  In this section and the next, profiles were calculated individually for each realization, and the median values of $\oO$ from each profile are plotted as the solid lines in these figures. The shaded areas around each line represent the range covered by the four realizations of each model.

All the simulations are characterized by a decline in the inner rotation as the system evolves. We point out, however, a notable feature in the time evolution of $\oO$ for the simulation vrQ001F05: in this model $\oO$ slightly increases during the early phases of the cluster evolution (for $t<5\trh$). We interpret this increase as a manifestation of the {\textbf{gravogyro instability}, a phenomenon in which angular momentum being transported outward causes the cluster to contract due to lack of centrifugal force and thus increases the angular speed near the centre of the cluster (see \citealt{inagaki1978} for the first theoretical investigation of this process in the context of fluid cylindrical shells, and \citealt{akiyama1989} and \citealt{ernst2007} for further discussion of this process, as an application to the interpretation of the evolution of N-body models with initial rotation).
Similar to what was discussed in the previous sections, the evolution of the inner angular velocity also appears to undergo a transition at the time of core collapse around $t\sim 15 \trh$. We emphasize that the inner rotation, although also undergoing a significant decrease during the cluster evolution, is still non-negligible after several tens of initial half-mass relaxation times and after the cluster has lost a large fraction of its initial mass. At the end of our simulations, $\oO \sim 1-2$; many real clusters are likely to be in a less advanced stage of their evolution and should therefore be  characterized by larger values of this ratio\footnote{Notice that the values of the angular velocity reported in these figures are those measured in the rotating frame of reference. Values in the non-rotating frame of reference are simply obtained as ($\omega_{\rm in}/\Omega)_{\rm nr-frame}=\oO+1$.}. 

Finally, in Fig. \ref{fig:twomethods} we compare the values of $\oO$ for the simulation vrQ001F05 obtained with the two different methods described above and we find them to be consistent with each other.  This further reinforces the potential for IFU data to greatly complement results from spectroscopic surveys focusing on individual stars.

We conclude this section by  pointing out that clusters starting with any of the initial conditions we have considered and currently at a similar evolutionary stage will be characterized by similar values of $\oO$; for clusters at different galactocentric distances, $R_g$, in a galaxy like the Milky Way with a flat rotation curve ($\Omega \propto 1/R_g$) this would result in a trend of $\omega_{\rm in}$ decreasing with $R_g$. A dispersion around this relation could arise as a result of a combination of effects due to differences in the initial cluster properties, in the cluster dynamical ages, and in the angle between the line of sight and the rotation axis. Interestingly, the results  of the observational study of \citet{fabricius2014} show some evidence of such a broad trend between  inner angular velocity and galactocentric distance.

\begin{figure}
	\includegraphics[width=3.3in,height=3.3in,page=1]{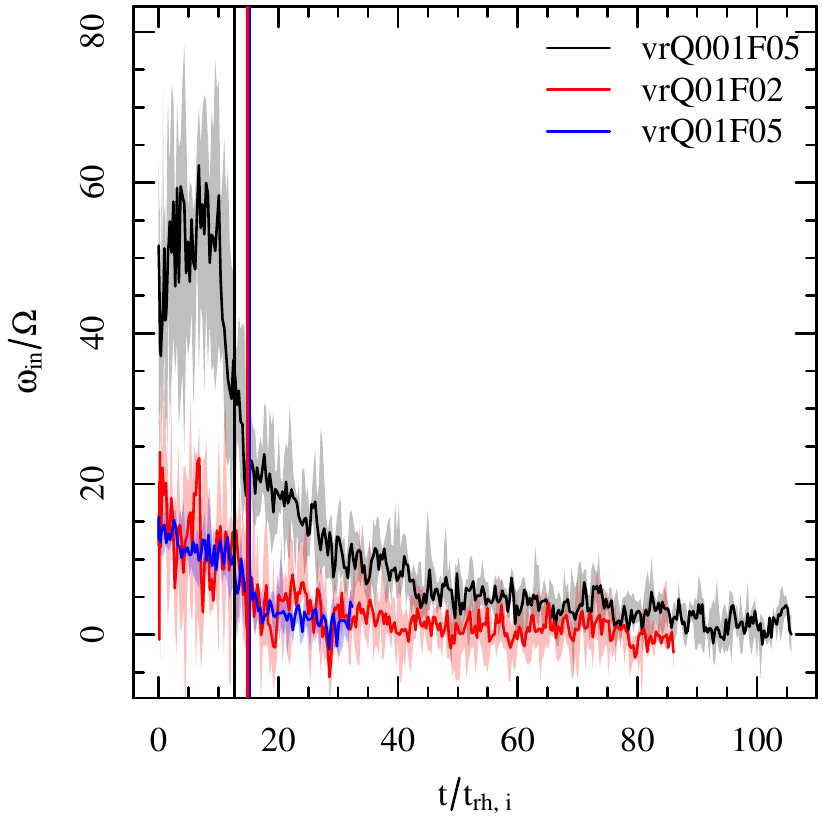}
    \caption{Evolution of the slope of the rotation curve between 0.1-0.5$\Rh$ for all the simulations.  Time is normalized to the initial half-mass relaxation time, while the slope is normalized to the cluster's orbital angular velocity.  The shaded areas represent the range covered by the four realizations for each model, and the vertical lines mark the median time of core collapse.}
    \label{fig:oOtrhi}
\end{figure}

\begin{figure}
	\includegraphics[width=3.3in,height=3.3in]{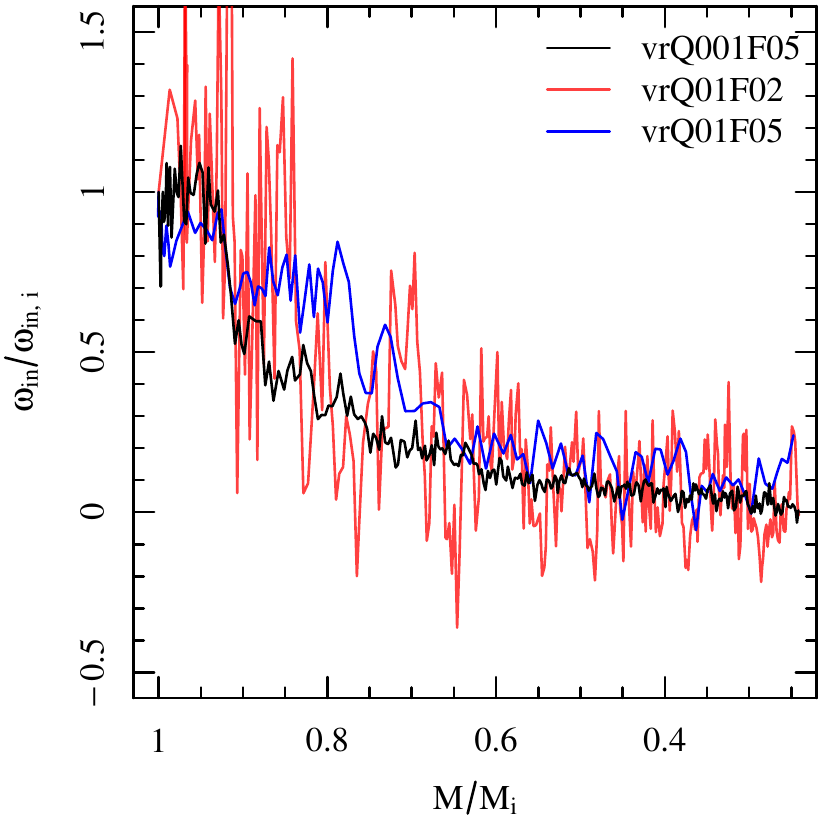}
    \caption{Evolution of the $\omega_{\rm in}$, normalized to its initial value as a function of fraction of initial mass remaining.}
    \label{fig:ominomini}
\end{figure}

\begin{figure}
	\includegraphics[width=3.3in,height=3.3in]{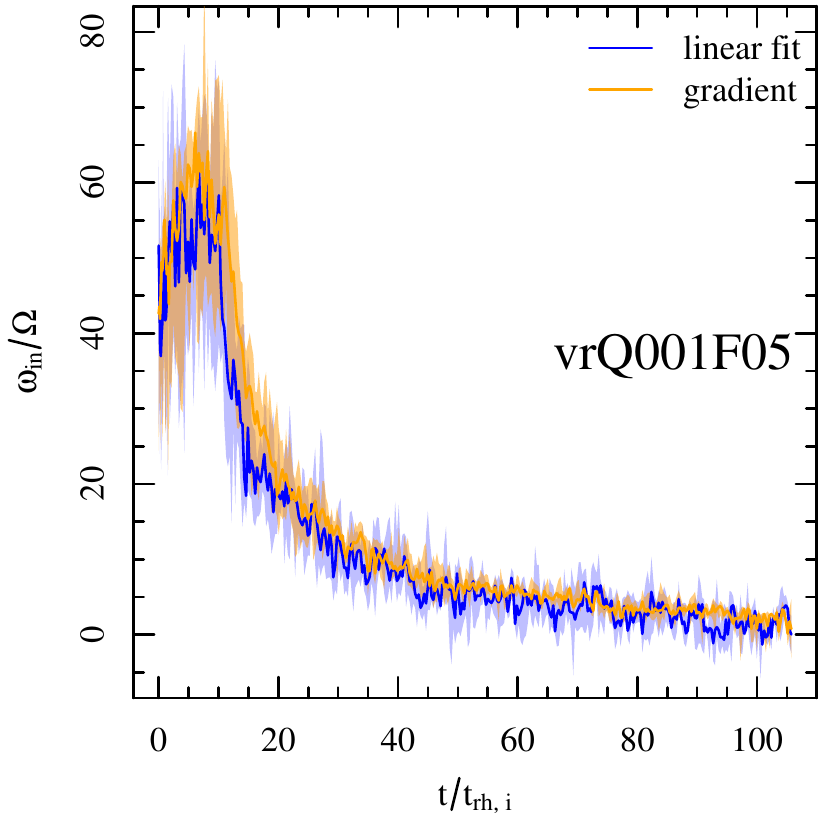}
    \caption{$\oO$ versus $t/\trh$ using the two different methods described in Section \ref{sec:innerevol}.  The shaded areas represent the range covered by the four realizations.}
    \label{fig:twomethods}
\end{figure}

\subsection{Evolution of the rotation in the outermost regions}
\label{sec:outerevol}

In \citet{tiongco2016b}, we showed that, as a result of the cooperation between the preferential loss of stars on prograde orbits and an intrinsic increase of the number of stars on retrograde orbits (which may be easily interpreted in light of an argument based on the conservation of angular momentum), initially non-rotating clusters acquire an approximately solid-body retrograde rotation in their outermost regions with angular velocity $\omega/\Omega \approx$ -0.5. 
In this section, we show that this effect also applies to the outer regions of the rotating models considered in this paper. 

Fig. \ref{fig:vphievol} shows that the systems we have explored here  have already acquired an outer retrograde rotation early during the violent relaxation phase \citep[see also][]{vesperini2014}.
In order to provide a more detailed characterization of the distribution of angular momentum in the outer regions, we have measured the slope of the rotation curve (calculated in the rotating frame of reference described in Section \ref{sec:methods}) for particles at radii $R > 0.5 \rj$. Hereafter we refer to this slope as $\oout$, and Figs \ref{fig:outer1} and \ref{fig:outer2} show the evolution of $\ooutO$ as a function of time (normalized to the initial half-mass relaxation time) and the fraction of the initial mass remaining in the cluster.  

The model with the larger initial ratio of the half-mass to Jacobi radius, $\rhrj$, (vrQ01F05) starts its evolution with a value $\ooutO$ approximately equal to -0.5 acquired during the violent relaxation phase and retains this value for the entire simulation. The two models with smaller initial values of $\rhrj$ (vrQ01F02 and vrQ001F05), emerge from the violent relaxation phase with $\ooutO$ approximately equal to -1, but after losing a small amount of mass their outer angular velocities converge to values close to  -0.5 for the rest of the simulation.

\section{Conclusions}
In this paper, we have presented the results of a survey of \Nbody simulations aimed at exploring the long-term evolution of rotation in star clusters evolving in the tidal field of the host galaxy.  All models acquire a differential rotation during their early evolution while they undergo a phase of violent relaxation; their rotation curves are characterized by a shape rising from the centre of the cluster, peaking at approximately 1-2 half-mass radii, then decreasing in the outer regions becoming slightly retrograde in the outermost regions.

We first discussed the evolution of the entire rotation curve of each of the models (Section \ref{sec:profevol}; Figs \ref{fig:vphievol}-\ref{fig:vsig0evol}) and then focused our attention on the properties of the rotation curve in the cluster innermost (Section \ref{sec:innerevol}) and outermost (Section \ref{sec:outerevol}) regions as well as on the strength and location of the peak of the rotational velocity (Section \ref{sec:peakevol}). 

Our simulations show that as a result of the transport of angular momentum from the inner to the outer regions and star escape carrying away angular momentum, the peak of the rotation curve and the slope of the rising inner part of the rotation curve decrease as the cluster evolves and lose stars. For all the models we have explored, the peak of the rotational velocity and the inner angular velocity normalized to their initial values have an approximately universal dependence on the fraction of the initial mass remaining in the cluster (see Figs \ref{fig:vpkm} and \ref{fig:ominomini}). We find that even after losing a significant fraction of their initial mass, clusters can retain some of their initial rotation.

We have quantified the relative strength of the ordered rotational motion to the random thermal motion by studying the evolution of the peak of the rotational velocity normalized to the central velocity dispersion $\vpksigmao$, as often done in observational studies, as well as the evolution of the peak of the ratio of the rotational velocity profile to the velocity dispersion profile, $\vphisigmap$. 
As the cluster evolves and loses part of its initial angular momentum, it becomes increasingly dominated by random motions, but even after several tens of initial relaxation times a cluster can still be characterized by non-negligible values of $\vpksigmao$ and $\vphisigmap$. In all cases our simulations show that the current rotation observed in clusters provides only a lower limit on the strength of initial rotation, and thus small current values of rotation should not be taken as an indication of the lack of importance of rotation in the past dynamical history of star clusters. 

We have explored the evolution of the location of the peak, $\rpk$, of the rotation curve over time and found  this to be located for most of the cluster evolution between $1-2 \Rh$ although it gradually evolves towards the cluster inner regions, and in the late phases of cluster evolution can be found at $\rpk/\Rh\sim 0.5-1$.

As for the rotation in the cluster outermost regions we have shown that all models evolve towards an approximately solid-body retrograde rotation with angular velocity approximately equal to half of the orbital angular velocity, in agreement with what found in our previous study for a set of initially non-rotating models  \citet{tiongco2016b}.

In a forthcoming paper, we will continue our investigation of the interplay between internal rotation and the effects of an external tidal field by further extending our investigation of the role of different structural and kinematical parameters on the evolution of the angular momentum of star clusters. Specifically, we will explore the long-term dynamical evolution of the kinematical and morphological properties, in the presence of a tidal perturbation, of a general class of rotating systems sampled from the family of distribution-function based models introduced by \citet{varri2012}.

\begin{figure}
	\includegraphics[width=3.3in,height=3.3in]{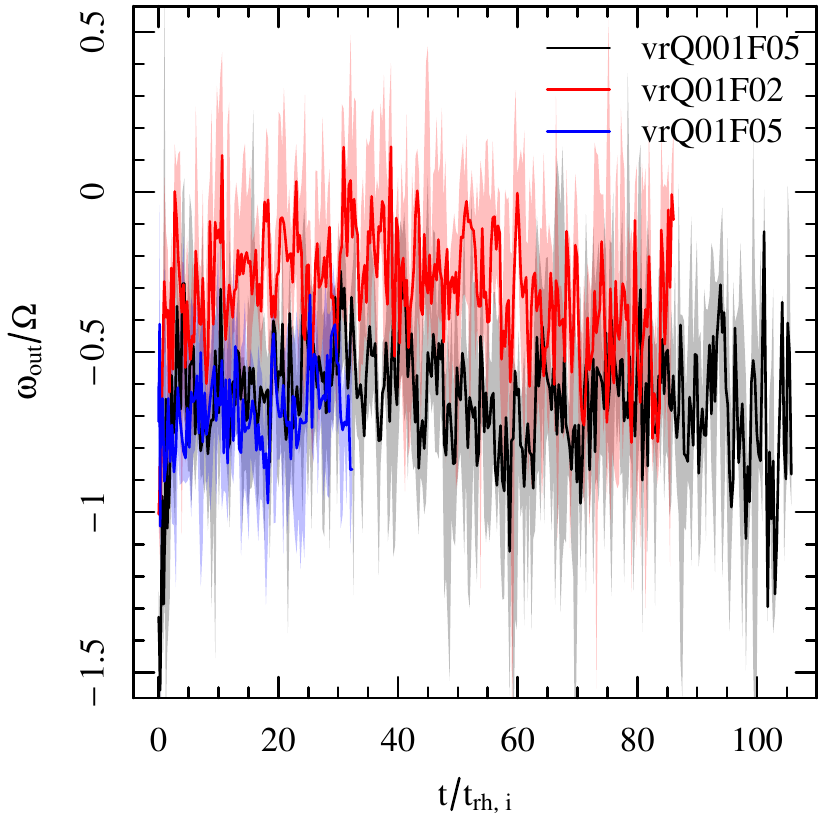}
    \caption{Time evolution of the slope of rotation curve in the outermost regions ($R > 0.5 \rj$) of the cluster for all model.  The shaded areas represent the range covered by the four realizations for each model.}
    \label{fig:outer1}
\end{figure}

\begin{figure}
	\includegraphics[width=3.3in,height=3.3in]{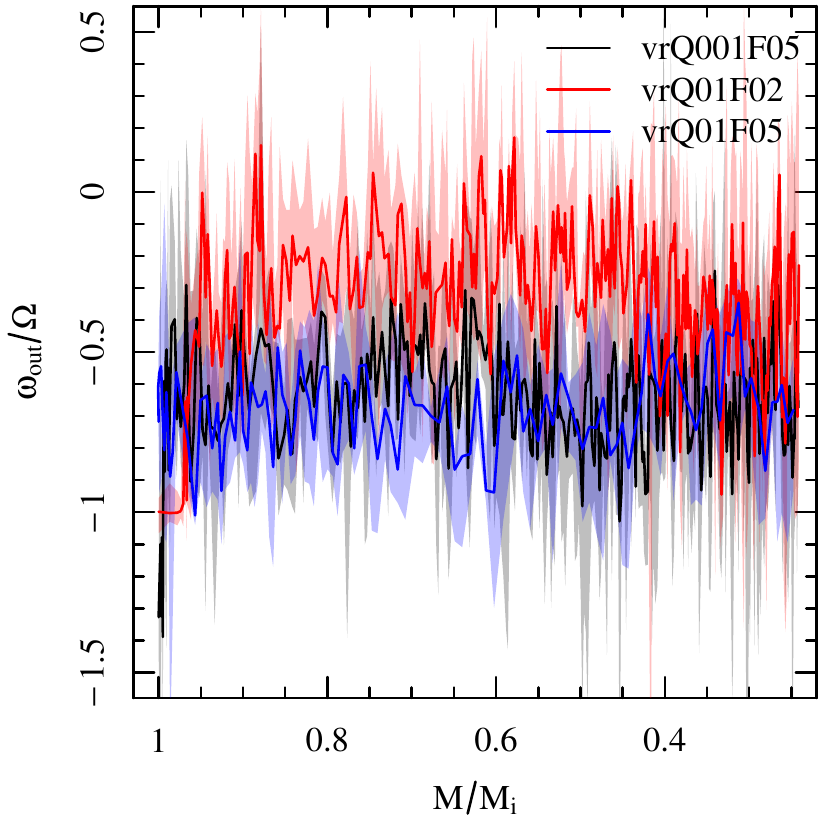}
    \caption{Similar to Fig. \ref{fig:outer1}, but as a function of fraction of mass remaining in the cluster.}
    \label{fig:outer2}
\end{figure}

\bibliographystyle{mnras}
\bibliography{references} 

\section*{Acknowledgements}

This research was supported in part by Lilly Endowment, Inc., through its support for the Indiana University Pervasive Technology Institute, and in part by the Indiana METACyt Initiative. The Indiana METACyt Initiative at IU is also supported in part by Lilly Endowment, Inc.  ALV is grateful to Douglas Heggie and Phil Breen for stimulating conversations, and to the EU Horizon 2020 program for support in form of a Marie Sklodowska-Curie Fellowship (MSCA-IF-EFRI 658088).

\bsp	
\label{lastpage}

\end{document}